
\input jytex.tex   
\typesize=10pt
\magnification=1200
\baselineskip=17truept
\hsize=6truein\vsize=8.5truein
\sectionnumstyle{blank}
\chapternumstyle{blank}
\chapternum=1
\sectionnum=1
\pagenum=0

\def\begintitle{\pagenumstyle{blank}\parindent=0pt\begin{narrow}[0.4in]}
\def\endtitle{\end{narrow}\newpage\pagenumstyle{arabic}}


\def\beginexercise{\vskip 20truept\parindent=0pt\begin{narrow}[10
truept]}
\def\endexercise{\vskip 10truept\end{narrow}}


\def\eql#1{\eqno\eqnlabel{#1}}
\def\ref{\reference}
\def\peq{\puteqn}
\def\pref{\putref}

\def\mgn{\marginnote}
\def\bex{\begin{exercise}}
\def\eex{\end{exercise}}

\def\mbox#1{{\leavevmode\hbox{#1}}}
\def\hspace#1{{\phantom{\mbox#1}}}

\def\al{\alpha}
\def\be{\beta}
\def\ga{\gamma}

\def\Ga{\Gamma}

\def\ep{\epsilon}

\def\ze{\zeta}

\def\De{\Delta}
\def\det{{\rm det\,}}

\def\Real{{\rm Re\,}}

\def\zf{$\zeta$--function}
\def\zfs{$\zeta$--functions}


\def\frac#1/#2{\leavevmode\kern.1em
\raise.5ex\hbox{\the\scriptfont0 #1}\kern-.1em/\kern-.15em
\lower.25ex\hbox{\the\scriptfont0 #2}}
\def\sfrac#1/#2{\leavevmode\kern.1em
\raise.5ex\hbox{\the\scriptscriptfont0 #1}\kern-.1em/\kern-.15em
\lower.25ex\hbox{\the\scriptscriptfont0 #2}}

\def\gtorder{\mathrel{\raise.3ex\hbox{$>$}\mkern-14mu
             \lower0.6ex\hbox{$\sim$}}}
\def\ltorder{\mathrel{\raise.3ex\hbox{$<$}\mkern-14mu
             \lower0.6ex\hbox{$\sim$}}}

\def\semidirprod{\rlap{\ss C}\raise1pt\hbox{$\mkern.75mu\times$}}
\def\for{\lower6pt\hbox{$\Big|$}}
\def\fish{\kern-.25em{\phantom{abcde}\over \phantom{abcde}}\kern-.25em}


\def\boxit#1{\vbox{\hrule\hbox{\vrule\kern3pt
        \vbox{\kern3pt#1\kern3pt}\kern3pt\vrule}\hrule}}
\def\dalemb#1#2{{\vbox{\hrule height .#2pt
        \hbox{\vrule width.#2pt height#1pt \kern#1pt
                \vrule width.#2pt}
        \hrule height.#2pt}}}


\def\noin{\noindent}

\def\comb#1#2{{\left(#1\atop#2\right)}}

\def\ie{{\it i.e. }}
\def\cf{{\it cf }}


  %

\def\sumstar#1{{\mathop{{\sum}^*_{#1}}}}

\def\3j#1#2#3#4#5#6{\left\lgroup\matrix{#1&#2&#3\cr#4&#5&#6\cr}
\right\rgroup}

\def\m?{\mgn{?}}


\def\aop#1#2#3{{\it Ann. Phys.} {\bf {#1}} (19{#2}) #3}

\def\cmp#1#2#3{{\it Comm. Math. Phys.} {\bf {#1}} (19{#2}) #3}
\def\cqg#1#2#3{{\it Class. Quant. Grav.} {\bf {#1}} (19{#2}) #3}

\def\jmp#1#2#3{{\it J. Math. Phys.} {\bf {#1}} (19{#2}) #3}
\def\jpa#1#2#3{{\it J. Phys.} {\bf A{#1}} (19{#2}) #3}

\def\np#1#2#3{{\it Nucl. Phys.} {\bf B{#1}} (19{#2}) #3}
\def\pl#1#2#3{{\it Phys. Lett.} {\bf {#1}} (19{#2}) #3}
\def\pm#1#2#3{{\it Phil.Mag.} {\bf {#1}} ({#2}) #3}

\def\pr#1#2#3{{\it Phys. Rev.} {\bf {#1}} (19{#2}) #3}
\def\prA#1#2#3{{\it Phys. Rev.} {\bf A{#1}} (19{#2}) #3}

\def\prD#1#2#3{{\it Phys. Rev.} {\bf D{#1}} (19{#2}) #3}

\def\prs#1#2#3{{\it Proc. Roy. Soc.} {\bf A{#1}} (19{#2}) #3}
\def\pcps#1#2#3{{\it Proc. Camb. Phil. Soc.} {\bf{#1}} (19{#2}) #3}

\def\dmj#1#2#3{{\it Duke Math. J.} {\bf {#1}} (19{#2}) #3}

\def\jdg#1#2#3{{\it J. Diff. Geom.} {\bf {#1}} (19{#2}) #3}
\def\jfa#1#2#3{{\it J. Func. Anal.} {\bf {#1}} (19{#2}) #3}

\def\ma#1#2#3{{\it Math. Ann.} {\bf {#1}} ({#2}) #3}
\def\mz#1#2#3{{\it Math. Zeit.} {\bf {#1}} ({#2}) #3}
\def\pams#1#2#3{{\it Proc. Am. Math. Soc.} {\bf {#1}} (19{#2}) #3}

\def\qjm#1#2#3{{\it Quart. J. Math.} {\bf {#1}} (19{#2}) #3}

\def\tams#1#2#3{{\it Trans. Am. Math. Soc.} {\bf {#1}} (19{#2}) #3}

\begin{title}
\vglue 20truept
\righttext {MUTP/95/12}
\righttext{hep-th/95}
\leftline{\today}
\vskip 100truept
\centertext {\Bigfonts \bf Oddball determinants}
\vskip 15truept
\centertext{J.S.Dowker\footnote{Dowker@a3.ph.man.ac.uk}}
\vskip 7truept
\centertext{\it Department of Theoretical Physics,\\
The University of Manchester, Manchester, England.}
\vskip 60truept
\centertext {Abstract}
\begin{narrow}
A simplified direct method is described for obtaining massless scalar
functional
determinants on the Euclidean ball. The case of odd \break dimensions is
explicitly discussed.

\end{narrow}
\vskip 5truept
\righttext {July 1995}
\vskip 75truept
\righttext{Typeset in \jyTeX}
\vfil
\end{title}
\pagenum=0
\section{\bf 1. Introduction}
In an earlier work, [\pref{Dow8}], a method of evaluating functional
determinants of the Laplacian, $D$, on the
even Euclidean ball was presented. In practice, the technique was
over-elaborate. It used the Watson-Kober summation result, an
exact formula involving Bessel functions, and thereby, since only an
asymptotic limit was required, unrequired information
was introduced, necessitating algebraic contortions in order to
reach the final goal. While some of these manipulations turned up
interesting identities and summations, they were strictly unnecessary
and somewhat of a luxury.  The object of the present, brief work is to
outline an improved method that emphasises the asymptotic behaviour more
immediately. The technique retains an element of the Watson-Kober approach
and generalises that of Moss [\pref{Moss}] employed by him {\it en
passant} while evaluating heat-kernel coefficients.
\section{\bf 2. Basic formulae}
Only a few of the starting formulae will be given here. Since our
previous work concerned even balls we concentrate, in the particular,
on odd $d$-balls. Hence the title of this paper.

The basic equation is (the $\approx$ sign means equal to the
mass-independent part of the large $m^2$ asymptotic limit)
$$
\ze'(0)=-\ln\det D\approx\lim_{m\to\infty}\ln\De(-m^2),
\eql{basicf}$$ where $\De$ is the Weierstrass product,
 $$
\De(-m^2)=\prod_{p,\al_p}(1+{m^2\over\al_p^2})\exp\sum_{k=1}^{[d/2]}
{1\over k}\big({-m^2\over\al_p^2}\big)^{k},
\eql{weier}$$
$\al_p^2$ being the eigenvalues of the relevant equation obtained by
setting
combinations of Bessel functions of order $p$ to zero. For simplicity the
equations are developed for Dirichlet conditions and then the
Mittag-Leffler theorem implies
$$
\De(-m^2)=\prod_p\big(p!2^pm^{-p}I_p(m^2)\big)\exp\sum_{k=1}^{[d/2]}
{1\over k}
\big({-m^2\over\al_p^2}\big)^{k}
\eql{weierb}$$
in terms of the modified Bessel function $I_p$.

The degeneracy, $N_p^{(d)}$, which is implied by the product in
(\peq{weierb}), is, for odd balls, an odd polynomial in $p$. Hence it is
enough to consider the quantity, [\pref{Dow8}],
$$\eqalign{
A_\nu=\ln\De(-m^2)&\sim\sum_{p=1/2}^\infty p^{2\nu+1}\bigg[p\ln{2p
\over p+\ep}+\ep-p
-{1\over2}\ln{\ep\over p}+\sum_{n=1}^\infty{T_n(t)\over \ep^n}\cr
&\hspace{***}-{\rm Ray}(m,p)+\int_0^\infty\bigg({1\over2}
-{1\over t}+{1\over e^t-1}\bigg){e^{-tp}\over t}\,dt\bigg],\cr}
\eql{logdet7}$$ where ${\rm Ray}(m,p)$ are the `Rayleigh' terms (\ie the
sums of inverse even powers of the eigenvalues). Olver's asymptotic form
of $I_p$ has been employed and an integral representation for $\ln p!$
substituted. The abbreviation $t=p/\ep$ with $\ep^2=p^2+m^2$ is frequently
used.

The central idea of simply discarding any $m$-dependent contributions to
(\peq{logdet7}) is the same as before, [\pref{Dow8}], and many of our
arguments are unchanged. Thus we immediately throw away the Rayleigh terms.
Also the discussion of the $T_n$ polynomial contribution and the integral
is identical. The major improvement is an accelerated treatment of the
remaining terms. To this end we set up the relevant asymptotic formulae.
\section{\bf 3. Asymptotic formulae}
Some standard results were given in our our earlier work and are
repeated here but in slightly generalised forms. The essential results are
$$
\sum_{p=1}^\infty {p^{2\nu+1}\over\ep^{2s}}\sim
\sum_{h=0}^\infty(1-h)\ldots(\nu-h)(m^2)^{1-h-s+\nu}{\Ga(s-1+h-\nu)
\over2\Ga(s)}{(-1)^hB_{2h}\over h!}
\eql{asym1}$$
and
$$
\sum_{p=1}^\infty{p^{2\nu}\over\ep^{2s}}
\sim{\Ga(\nu+1/2)\Ga\big(s-\nu-1/2\big)
\over2\Ga(s)}m^{1-2s+2\nu},\quad\nu>0.
\eql{Moss1}$$

{}From these formulae it is easy to derive
$$\eqalign{
&\sum_{p=1/2}^\infty{p^{2\nu+1}\over\ep^{2s}}\cr
&\sim
\sum_{h=0}^\infty(1-h)\ldots(\nu-h)\big(2^{1-2h}-1\big)
(m^2)^{1-h-s+\nu}{\Ga(s-1+h-\nu)
\over2\Ga(s)}{(-1)^hB_{2h}\over h!}\cr}
\eql{asym2}$$
and
$$
\sum_{p=1/2}^\infty{p^{2\nu}\over\ep^{2s}}
\sim{\Ga(\nu+1/2)\Ga\big(s-\nu-1/2\big)
\over2\Ga(s)}m^{1-2s+2\nu},\quad\nu>0,
\eql{Moss2}$$
which are more suitable here.
\section{\bf 4. Evaluation of the determinant}
Now we look at the first three terms in the bracket in (\peq{logdet7}).
Since the entire expression (\peq{logdet7}) converges as a summation, it is
possible to render any part finite by systematic addition and subtraction
of appropriate terms. This is done by the removal of enough of the Taylor
series to regularise the summation, a process denoted by a star. For example,
$$\eqalign{
\sum_{p=1/2}^\infty p^{2\nu+1}\ln\big({\ep\over p}\big)\to&
\sumstar{p=1/2}\, p^{2\nu+1}\ln\big({\ep\over p}\big)\cr
&\equiv
\sum_{p=1/2}^\infty p^{2\nu+1}\bigg(\ln\big({\ep\over p}\big)-{1\over2}
\sum_{k=1}^{\nu+1}(-1)^{k+1}{m^{2k}\over p^{2k}}\bigg). \cr}
\eql{term1r}$$

Consider now the same procedure applied to the quantities in
(\peq{asym2}) and (\peq{Moss2})
$$\sumstar{p}{p^N\over\ep^{2s}}=
\sum_{p}^\infty p^N\bigg[{1\over\ep^{2s}}-{1\over p^{2s}}-
\sum_{k=1}^M\comb{-s}k{m^{2k}\over p^{2k+2s}}\bigg]
\eql{watmoss}$$
whose derivative at $s=0$ gives minus twice (\peq{term1r}) when
$N=2\nu+1$ and $M=\nu+1$.

Using (\peq{asym2}) on the first term on the right-hand side in
(\peq{watmoss}), performing the summations on the others to give
Riemann \zfs\ and then differentiating, one obtains the
asymptotic limit of (\peq{term1r}). The only term independent of $m^2$
is clearly
$$\ze_R'(-2\nu-1,1/2)=(2^{-2\nu-1}-1)\ze_R'(-2\nu-1)+2^{-2\nu-1}
\ze_R(-2\nu-1).
\eql{term1f}$$

We next turn to the term which caused most problems in our earlier
calculation,
$$\eqalign{
\sum_{p=1/2} p^{2\nu+2}\ln{2p\over p+\ep}&=
\sum_{p=1/2}^\infty p^{2\nu+2}\bigg[\ln{2p\over\ep}+
\sum_{k=1}^\infty{(-1)^k\over k}{p^k\over\ep^k}\bigg]\cr
&=\sum_{p=1/2}^\infty p^{2\nu+2}\bigg[\ln{p\over\ep}+
\sum_{k=1}^\infty{(-1)^k\over k}\bigg({p^k\over\ep^k}-1\bigg)\bigg],\cr}
\eql{term2}$$
after an expansion in $p/\ep$.

The regularised version of this can be formally rearranged,
$$
\sumstar{p=1/2} p^{2\nu+2}\ln{2p\over p+\ep}=
\sumstar{p=1/2}p^{2\nu+2}\ln{p\over\ep}+
\sum_{k=1}^\infty{(-1)^k\over k}\sumstar{p=1/2}{p^{2\nu+2+k}\over\ep^k}
\eql{term2reg}$$
so that the asymptotic formulae can again be used. The logarithm has
effectively already been
considered and (\peq{watmoss}) can be applied to the last summation setting
$s=k/2$ and $N=2\nu+2+k$. The mass-independent term, for all $k$ (even
and odd), is easily seen to equal $\ze_R(-2\nu-2,1/2)$ which is zero,
whence the total is
$$-\ze_R'(-2\nu-2,1/2)=-(2^{-2\nu-2}-1)\,
\ze_R'(-2\nu-2).
\eql{term2f}$$

Incidentally, the corresponding calculation for even balls has the factor
$p^{2\nu+1}$ in (\peq{term2}) and a summation over integers. It rapidly
gives the answer found in [\pref{Dow8}],
$$
-\ze_R'(-2\nu-1)-\ln2\,\ze_R(-2\nu-1),
$$
without the complicated manipulations of MacDonald functions that occupied
us so excessively there.

The asymptotic behaviour of the remaining term in (\peq{logdet7})
follows from (\peq{watmoss}) with $N=2\nu+1$ and $s\to-1/2$ and is
$$
\sum_{p}^\infty p^{2\nu+1}\big(\ep-p\big)\to
\sumstar{p}p^{2\nu+1}(\ep-p)=
\sum_{p}^\infty p^{2\nu+1}\bigg(\ep-p-{1\over2}{m^2\over p}\bigg),
$$
giving a mass-independent part of $-\ze_R(-2\nu-2,1/2)$, which vanishes.
\mgn{FACTOR?}

As remarked, the general discussion of the remaining terms in
(\peq{logdet7}) is exactly that of our previous work. This time
there is no escape from the algebra but since the details are a little
different from those earlier ones, and slightly easier, some elaboration
is not out of place here.

As explained in [\pref{Dow8}], by adding and subtracting terms, and
discarding obviously mass-dependent convergent contributions (in the
limit)
from the $T_n$ polynomials, one arrives at
$$\eqalign{
\sum_{p=1/2}^\infty p^{2\nu+1}\bigg[\sum_{n=1}^{2\nu+2}\bigg(&{T_n(t)
\over\ep^n}-{T_n(1)\over p^n}\bigg)\cr
&+\int_0^\infty\!\!\bigg({1\over2}
-{1\over\tau}+\!\sum_{k=1}^{\nu+1}(-1)^kB_{2k}{\tau^{2k-1}\over(2k)!}
+{1\over e^\tau-1}\bigg){e^{-\tau p}\over \tau }\,d\tau \bigg]\cr}
\eql{rem2}$$
for these remaining terms, the parts of which are now considered. First
$$
\sum_{p=1}^\infty p^{2\nu+2}\sum_{n=1}^{2\nu+1}\bigg({T_n(t)\over\ep^n}-
{T_n(1)\over p^n}\bigg).
\eql{pol1}$$
Repeating our previous analysis we set
$$
T_n(t)=T_n(1)+T'_n(t)
$$
and get for (\peq{pol1})
$$
\sum_{n=1,3,\ldots}^{2\nu+1}\,\,\sum_{p=1/2}^\infty T_n(1) p^{2\nu+1}
\bigg({1\over\ep^n}-{1\over p^n}\bigg)
+\sum_{n=1}^{2\nu+2}\sum_{p=1/2}^\infty
{p^{2\nu+1}\over\ep^n}T'_n(t)
\eql{pol2}$$
noting that $T_n(1)$ vanishes for $n$ even.

We apply (\peq{watmoss}) with $N=2\nu+1$ and $s=n/2$ to the first summation.
Doing the summations in (\peq{watmoss}), we note that none of the  Riemann
\zf\ arguments can equal $1$, because $2s$ is odd. For the same reason,
there is no mass dependence in the asymptotic limit, (\peq{asym2}), of the
first term on the right-hand side of (\peq{watmoss}). Since the last summation
in (\peq{watmoss}) can be discarded as mass-dependent, this just leaves the
contribution of the second term, which is
$$
-\sum_{n=1,3,\ldots}^{2\nu+1}T_n(1)\,\ze_R(n-2\nu-1,1/2)
$$
and vanishes because $n$ is odd.

In the second summation in (\peq{pol2}) only the convergent term
$n=2\nu+2$ yields a mass-independent part to the asymptotic limit according
to (\peq{asym2}) and manipulations similar to those outlined in our earlier
work produce the value
$$
\int_0^1 t^{2\nu+1}T''_{2\nu+2}(t)\,dt
\eql{polytot}$$
for this constant contribution where $T_n'(t)= (1-t^2)T_n''(t)$.

Turning to the integral in ({\peq{rem2}), the sum over $p$ is
performed and a regularisation introduced to give
$$
\lim_{s\to0}\int_0^\infty\bigg({1\over2}
-{1\over \tau }+\sum_{k=1}^{\nu+1}(-1)^kB_{2k}{\tau ^{2k-1}\over(2k)!}
+{1\over e^\tau -1}\bigg)\tau ^{s-1}(-1)^{2\nu+1}{d^{2\nu+1}\over d\tau
^{2\nu+1}}{e^{\tau /2}\over e^\tau -1}\,d\tau .
\eql{int3}$$

The power terms in (\peq{int3}) are dealt with using the formula
$$
(-1)^j\int_0^\infty \tau ^\mu \tau ^{s-1}{d^j\over d\tau ^j}{e^{\tau /2}
\over e^\tau -1}\,d\tau=
\Ga(\mu+s)\ze_R(\mu-j+s,1/2),
\eql{powers1}$$
while the remaining part,
$$
(-1)^{2\nu+1}\lim_{s\to0}\int_0^\infty{1\over e^\tau -1}\tau ^{s-1}
{d^{2\nu+1}\over d\tau ^{2\nu+1}}{e^{\tau /2}\over e^\tau -1}\,d\tau ,
\eql{dint}$$
is best treated by writing
$$
(-1)^j{d^j\over d\tau ^j}{e^{\tau /2}\over e^\tau -1}=\sum_{l=1}^{j+1}
D_l^{(j)}{e^{\tau /2}\over (e^\tau -1)^l}
$$with the recursion\mgn{EXPCOEFS.MTH}
$$\eqalign{
&D_l^{(j)}=(l-1/2) D_l^{(j-1)}+(l-1)D_{l-1}^{(j-1)},\quad 2\le l\le j,\cr
&D_{j+1}^{(j)}=j!\cr
&D_1^{(j)}={1\over2^j}.\cr}
\eql{recurs2}$$
A special value is $D^{(j)}_j=jj!/2$.

Expression (\peq{dint}) can be rewritten as
$$
\lim_{s\to0}\sum_{l=1}^{2\nu+2}D_l^{(2\nu+1)}\,\Ga(s)\,\ze_{l+1}
\big(s,l+1/2\big)
\eql{dint2}$$
in terms of the Barnes \zf,
$$\eqalign{
\ze_r(s,a)&={i\Ga(1-s)\over2\pi}\int_L{e^{z(r-a)}(-z)^{s-1}\over
(e^z-1)^r}\,dz\cr
&=\sum_{n=0}^\infty\comb{n+r-1}{r-1}{1\over{(a+n)}^s}\,,\quad \Real s>r.
\cr}
\eql{hemibarnes}$$ In the present case $r=l+1$, $a=l+1/2$ and the lower
limit can be adjusted so that
$$
\ze_{l+1}(s,l+1/2)=\sum_{n=0}^\infty\comb nl{1\over (n+1/2)^s}.
$$
As usual the numerator is expanded in powers of $(n+1/2)$ using
Stirling numbers,
$$(n+a)(n+a-1)\ldots(n+a-b+1)=\sum_{k=0}^bT^{(k)}(a,b)\,(n+1/2)^k,
$$
to give the Barnes function as a series of standard Riemann-Hurwitz \zfs,
$$
\ze_{l+1}(s,l+1/2)={1\over l!}\sum_{k=0}^lT^{(k)}(0,l)\,\ze_R(s-k,1/2),
\eql{barnexp}$$
which are the natural quantities in the odd case.

Putting the two parts of the integral together gives
$$\eqalign{
\lim_{s\to0}\bigg({1\over2}\Ga(s)&\ze_R(s-2\nu-1,1/2)
-\Ga(s-1)\ze_R(s-2\nu-2,1/2)\cr
&-\sum_{k=1}^{\nu+1}(-1)^{k+1}{B_{2k}\over(2k)!}\Ga(s+2k-1)\ze_R(s+2k
-2\nu-2,1/2)\cr
&\hspace{*****}+\Ga(s)\sum_{l=1}^{2\nu+2}\sum_{k=0}^l\,D_l^{(2\nu+1)}
\,{T^{(k)}(0,l)\over l!}\,\ze_R(s-k,1/2) \bigg).\cr}
\eql{total2}$$

The cancellation of the individual divergences is a check of the analysis
and implies the identity
$$
{1\over2}\ze_R(-2\nu-1,1/2)
+\sum_{l=1}^{2\nu+2}
\sum_{k=0}^l\,D_l^{(2\nu+1)}\,{T^{(k)}(0,l)\over l!}\,\ze_R(-k,1/2)=0,
\eql{polcanc}$$
where $k$ must be odd.

The finite remainder in (\peq{total2}) is
$$\eqalign{
{1\over2}\ze_R'(&-2\nu-1,1/2)-{1\over2}\ga\ze_R(-2\nu-1)
+\ze_R'(-2\nu-2,1/2)\cr
&\hspace{*****}+\sum_{l=1}^{2\nu+2}\sum_{k=0}^l\,D_l^{(2\nu+1)}\,
{T^{(k)}(0,l)\over l!}\,\big(\ze'_R(-k,1/2)-\ga\ze_R(-k,1/2)\big)\cr}
$$
which can be reduced using (\peq{polcanc}) leaving,
$$
{1\over2}\ze_R'(-2\nu-1,1/2)+\ze_R'(-2\nu-2,1/2)
+\sum_{l=1}^{2\nu+2}\sum_{k=0}^l\,D_l^{(2\nu+1)}\,{T^{(k)}(0,l)\over l!}
\,\ze'_R(-k,1/2).
\eql{total3}$$

Adding the contribution (\peq{term2f}) from  (\peq{term2reg}),
cancels the second term in (\peq{total3}) while that, (\peq{term1f}), from
(\peq{term1r}) cancels the first one. These cancellations suggest that
there is a yet more efficient way of organising the asymptotic limits.

The order of the summations in the remaining term in (\peq{total3}) is
reversed by writing
$$
\sum_{l=1}^{2\nu+2}\sum_{k=0}^l\,D_l^{(2\nu+1)}\,{T^{(k)}(0,l)\over l!}
\,\ze'_R(-k,1/2)=\sum_{k=0}^{2\nu+2}N_k^{(2\nu+1)}\ze_R'(-k,1/2)
$$
where the vector of coefficients is\mgn{STIR.MTH}
$$
N_k^{(2\nu+1)}\equiv \sum_{l=k}^{2\nu+2}\,D_l^{(2\nu+1)}\,
{T^{(k)}(0,l)\over l!}
$$
with $D^{(2\nu+1)}_0=0$.

Including the contribution (\peq{polytot}) yields the final expression for
(\peq{logdet7}),
$$
A_\nu=\sum_{k=0}^{2\nu+2}N_k^{(2\nu+1)}\,\ze'_R(-k,1/2)
+\int_0^1t^{2\nu+1}T''_{2\nu+2}(t)\,dt.
\eql{oddfin}$$

The complete determinant for any odd ball is obtained by compounding
(\peq{oddfin}) with the degeneracy expressed as a polynomial in $p$.
For the 3-ball the degeneracy is $2p$ and the answer is
$$\ze_3'(0)=-{3\over32}+\ze_R'(-2,1/2)-\ze_R'(-1,1/2)+{1\over4}
\ze_R'(0,1/2)\approx-0.21139
\eql{det3ball}$$
in agreement, after reverting to ordinary Riemann \zfs, with our earlier
calculation [\pref{DandA}] and with that of Bordag {\it et al}\  [\pref{BGKE}],
who use another direct method.

For the 5-ball the degeneracy is $p^3/3-p/12$ and
$$\eqalign{
\ze_5'(0)={47\over9216}&+{1\over12}\ze_R'(-4,1/2)-{1\over6}
\ze_R'(-3,1/2)+{1\over24}\ze_R'(-2,1/2)\cr
&+{1\over24}\ze_R'(-1,1/2)-{1\over64}
\ze_R'(0,1/2)\,\approx\,0.01375,\cr}
\eql{det5ball}$$
which again agrees with [\pref{BGKE}] after rearrangement.

For the general $d$-ball it is straightforward to write a symbolic manipulation
programme that performs all the operations automatically. Number enthusiasts
might like to know that the 7-ball value is
$$\eqalign{
\ze'_7(0)=-&{11831502329\over19263179980800}+{1\over360}\ze_R'(-6,1/2)
-{1\over120}\ze_R'(-5,1/2)\cr
&-{1\over288}\ze_R'(-4,1/2)
+{1\over48}\ze_R'(-3,1/2)-{41\over5760}\ze_R'(-2,1/2)\cr
&-{3\over640}\ze_R'(-1,1/2)+{1\over512}\ze_R'(0,1/2)\,\approx\,-0.001751.\cr}
\eql{det7ball}$$
\section{\bf 5. Robin boundary conditions.}

The essential equation in the Robin case is (\cf [\pref{Dow8}])
$$\eqalign{
A_\nu(\be)=\ln&\De\big(-m^2\big)\cr
&\sim\sumstar{p}p^{2\nu+1}\bigg(p\ln{2p\over p+\ep}+\ep-p
+{1\over2}\ln{\ep\over p}-\ln(1+\be/p)\bigg)\cr
&+\sum_{p=1}^\infty p^{2\nu+1}\bigg[\sum_{n=1}^{2\nu+2}{R_n(\be,t)
-R_n(\be,1)
\over\ep^n}+\sum_{n=1}^{2\nu+2}R_n(\be,1)\bigg({1\over\ep^n}-
{1\over p^n}\bigg)\cr
&\hspace{*}+\int_0^\infty\bigg({1\over2}
-{1\over \tau }-\!\sum_{k=1}^{\nu+1}(-1)^{k+1}B_{2k}{\tau ^{2k-1}\over(2k)!}
+{1\over e^\tau -1}\bigg){e^{-\tau p}\over \tau }\,d\tau \bigg]\cr}
\eql{logdetr2}$$
where the $R_n(\be,t)$ are polynomials arising from a cumulant
expansion of Olver's asymptotic series of the appropriate Robin
combination of Bessel functions.

Most of the analysis is the same as for Dirichet
conditions. We only note that
$$\eqalign{
\sumstar{p=1/2}p^{2\nu+1}\ln(1+\be/p)&={\ga\over2(\nu+1)}\be^{2\nu+2}
+\!\!\int_0^\be\!\!\be^{2\nu+1}\big(2\psi(1\!+\!2\be)\!-\!\psi(1\!+\!\be)
\big)\,d\be\cr
&={\ga+2\ln2\over2(\nu+1)}\be^{2\nu+2}
+\int_0^\be\be^{2\nu+1}\psi(1/2+\be)\,d\be.\cr}
\eql{exterm}$$

It is necessary, however, to be careful when working out the contribution of
the term
$$
\sum_{n=1}^{2\nu+2}R_n(\be,1)\,\sumstar{p}p^{2\nu+1}\bigg({1\over\ep^n}-
{1\over p^n}\bigg)
\eql{ext}$$ because $R_n(\be,1)=-(-1)^n\be^n/n$ for even $n$ and there are
extra mass-independent terms when $\nu>0$ coming from the application of
(\peq{watmoss}) and (\peq{asym2}).

Combining the different contributions produces the final expression for
(\peq{logdetr2}),
$$\eqalign{
A_\nu(\be)=\ze_R'&(-2\nu-1)+\sum_{k=0}^{2\nu+2}N_k^{(2\nu+1)}\,
\ze'_R(-k,1/2)\cr
&+{\be^{2\nu+2}\over4(\nu+1)}\sum_1^\nu{1\over k}
-\sum_{L=0}^{\nu-1} \be^{2\nu-2L}{\big(2^{-2L-1}-1\big)B_{2+2L}
\over4(\nu-L)(L+1)}\cr
&-\int_0^\be\be^{2\nu+1}
\psi(1/2+\be)\,d\be+\int_0^1t^{2\nu+1}R''_{2\nu+2}(\be,t)\,dt.\cr}
\eql{oddfinr}$$
Specifically, for the 3-ball,
$$\eqalign{
\ze_3'(0)={3\over32}+&\ze_R'(-2,1/2)+{1\over2}\ze_R'(-1,1/2)+{1\over4}
\ze_R'(0,1/2) +{\be\over2}\cr
&-\int_0^\be \be\psi(1/2+\be)\,d\be,\cr}
\eql{det3ballr}$$
and for the 5-ball,
$$\eqalign{
\ze_5'(0)=-&{61\over46080}+{1\over12}\ze_R'(-4,1/2)+{1\over6}
\ze_R'(-3,1/2)+{1\over24}\ze_R'(-2,1/2)\cr
&-{1\over24}\ze_R'(-1,1/2)-{1\over64}\ze_R'(0,1/2)+{1\over24}\be^4
+{1\over24}\be^3+{1\over48}\be^2 -{11\over576}\be\cr
&+{1\over12}\int_0^\be\be(1-4\be^2)\psi(1/2+\be)\,d\be,\cr}
\eql{det5ballr}$$
which are again ultimately in agreement with [\pref{BGKE}].

The fact that the coefficients of the \zf\ derivatives are the same as the
Dirichlet ones, up to signs, is a consequence of the special value
$N^{(2\nu+1)}_{2\nu+1}=-1/2$.
\section{\bf 6. An identity}
Instead of using (\peq{asym2}) directly to evaluate the mass-independent
extra terms that come from (\peq{ext}), it is possible to expand
$p^{2\nu}=(\ep^2-p^2)^\nu$ binomially and use (\peq{asym2}) for $\nu=0$ on each
term. This is equivalent to the known identity
$$
B(\nu-L,L+1)=
\sum_{q=0}^L{(-1)^L\over\nu-q}\comb{L}{q}
$$
for the Euler $\be$-function.
\section{\bf 7. Comments}
Although the odd ball results have been emphasised, the method applies
equally well to the even case. One notes the oscillating variation of the
determinants with dimension.

It is relatively easy to extend the
results to the spinor Laplacian. One can also treat certain portions of the
ball.
\section{\bf Acknowledgment}
I would like to thank Emilio Elizalde for helpful correspondance.

\vskip 10truept
\noin{\bf{References}}
\vskip 5truept
\begin{putreferences}
\ref{Rayleigh}{Lord Rayleigh{\it Theory of Sound} vols.I and II,
MacMillan, London, 1877,78.}
\ref{KCD}{G.Kennedy, R.Critchley and J.S.Dowker \aop{125}{80}{346}.}
\ref{Donnelly} {H.Donnelly \ma{224}{1976}161.}
\ref{Fur2}{D.V.Fursaev {\sl Spectral geometry and one-loop divergences on
manifolds with conical singularities}, JINR preprint DSF-13/94,
hep-th/9405143.}
\ref{HandE}{S.W.Hawking and G.F.R.Ellis {\sl The large scale structure of
space-time} Cambridge University Press, 1973.}
\ref{DandK}{J.S.Dowker and G.Kennedy \jpa{11}{78}{895}.}
\ref{ChandD}{Peter Chang and J.S.Dowker \np{395}{93}{407}.}
\ref{FandM}{D.V.Fursaev and G.Miele \pr{D49}{94}{987}.}
\ref{Dowkerccs}{J.S.Dowker \cqg{4}{87}{L157}.}
\ref{BandH}{J.Br\"uning and E.Heintze \dmj{51}{84}{959}.}
\ref{Cheeger}{J.Cheeger \jdg{18}{83}{575}.}
\ref{SandW}{K.Stewartson and R.T.Waechter \pcps{69}{71}{353}.}
\ref{CandJ}{H.S.Carslaw and J.C.Jaeger {\it The conduction of heat
in solids} Oxford, The Clarendon Press, 1959.}
\ref{BandH}{H.P.Baltes and E.M.Hilf {\it Spectra of finite systems}.}
\ref{Epstein}{P.Epstein \ma{56}{1903}{615}.}
\ref{Kennedy1}{G.Kennedy \pr{D23}{81}{2884}.}
\ref{Kennedy2}{G.Kennedy PhD thesis, Manchester (1978).}
\ref{Kennedy3}{G.Kennedy \jpa{11}{78}{L173}.}
\ref{Luscher}{M.L\"uscher, K.Symanzik and P.Weiss \np {173}{80}{365}.}
\ref{Polyakov}{A.M.Polyakov \pl {103}{81}{207}.}
\ref{Bukhb}{L.Bukhbinder, V.P.Gusynin and P.I.Fomin {\it Sov. J. Nucl.
 Phys.} {\bf 44} (1986) 534.}
\ref{Alvarez}{O.Alvarez \np {216}{83}{125}.}
\ref{DandS}{J.S.Dowker and J.P.Schofield \jmp{31}{90}{808}.}
\ref{Dow1}{J.S.Dowker \cmp{162}{94}{633}.}
\ref{Dow2}{J.S.Dowker \cqg{11}{94}{557}.}
\ref{Dow3}{J.S.Dowker \jmp{35}{94}{4989}; erratum {\it ibid}, Feb.1995.}
\ref{Dow5}{J.S.Dowker {\it Heat-kernels and polytopes} To be published}
\ref{Dow6}{J.S.Dowker \pr{D50}{94}{6369}.}
\ref{Dow7}{J.S.Dowker \pr{D39}{89}{1235}.}
\ref{Dow8}{J.S.Dowker {\it Robin conditions on the Euclidean ball}
hep-th/9506042.}
\ref{BandG}{P.B.Gilkey and T.P.Branson \tams{344}{94}{479}.}
\ref{Schofield}{J.P.Schofield Ph.D.thesis, University of Manchester,
(1991).}
\ref{Barnesa}{E.W.Barnes {\it Trans. Camb. Phil. Soc.} {\bf 19} (1903)
374.}
\ref{BandG2}{T.P.Branson and P.B.Gilkey {\it Comm. Partial Diff. Equations}
{\bf 15} (1990) 245.}
\ref{Pathria}{R.K.Pathria {\it Suppl.Nuovo Cim.} {\bf 4} (1966) 276.}
\ref{Baltes}{H.P.Baltes \prA{6}{72}{2252}.}
\ref{Spivak}{M.Spivak {\it Differential Geometry} vols III, IV, Publish
or Perish, Boston, 1975.}
\ref{Eisenhart}{L.P.Eisenhart {\it Differential Geometry}, Princeton
University Press, Princeton, 1926.}
\ref{Moss}{I.Moss \cqg{6}{89}{659}.}
\ref{Barv}{A.O.Barvinsky, Yu.A.Kamenshchik and I.P.Karmazin \aop {219}
{92}{201}.}
\ref{Kam}{Yu.A.Kamenshchik and I.V.Mishakov {\it Int. J. Mod. Phys.}
{\bf A7} (1992) 3265.}
\ref{DandE}{P.D.D'Eath and G.V.M.Esposito \prD{43}{91}{3234}.}
\ref{Rich}{K.Richardson \jfa{122}{94}{52}.}
\ref{Osgood}{B.Osgood, R.Phillips and P.Sarnak \jfa{80}{88}{148}.}
\ref{BCY}{T.P.Branson, S.-Y. A.Chang and P.C.Yang \cmp{149}{92}{241}.}
\ref{Vass}{D.V.Vassilevich.{\it Vector fields on a disk with mixed
boundary conditions} gr-qc /9404052.}
\ref{MandP}{I.Moss and S.Poletti \pl{B333}{94}{326}.}
\ref{Kam2}{G.Esposito, A.Y.Kamenshchik, I.V.Mishakov and G.Pollifrone
\prD{50}{94}{6329}.}
\ref{Aurell1}{E.Aurell and P.Salomonson \cmp{165}{94}{233}.}
\ref{Aurell2}{E.Aurell and P.Salomonson {\it Further results on functional
determinants of laplacians on simplicial complexes} hep-th/9405140.}
\ref{BandO}{T.P.Branson and B.\O rsted \pams{113}{91}{669}.}
\ref{Elizalde1}{E.Elizalde, \jmp{35}{94}{3308}.}
\ref{BandK}{M.Bordag and K.Kirsten {\it Heat-kernel coefficients of
the Laplace operator on the 3-dimensional ball} hep-th/9501064.}
\ref{Waechter}{R.T.Waechter \pcps{72}{72}{439}.}
\ref{GRV}{S.Guraswamy, S.G.Rajeev and P.Vitale {\it O(N) sigma-model as
a three dimensional conformal field theory}, Rochester preprint UR-1357.}
\ref{CandC}{A.Capelli and A.Costa \np {314}{89}{707}.}
\ref{IandZ}{C.Itzykson and J.-B.Zuber \np{275}{86}{580}.}
\ref{BandH}{M.V.Berry and C.J.Howls \prs {447}{94}{527}.}
\ref{DandW}{A.Dettki and A.Wipf \np{377}{92}{252}.}
\ref{Weisbergerb} {W.I.Weisberger \cmp{112}{87}{633}.}
\ref{Voros}{A.Voros \cmp{110}{87}{110}.}
\ref{Pockels}{F.Pockels {\it \"Uber die partielle Differentialgleichung
$\Delta u+k^2u=0$}, B.G.Teubner, Leipzig 1891.}
\ref{Kober}{H.Kober \mz{39}{1935}{609}.}
\ref{Watson2}{G.N.Watson \qjm{2}{31}{300}.}
\ref{DandC1}{J.S.Dowker and R.Critchley \prD {13}{76}{3224}.}
\ref{Lamb}{H.Lamb \pm{15}{1884}{270}.}
\ref{EandR}{E.Elizalde and A.Romeo International J. of Math. and Phys.
{\bf13} (1994) 453}
\ref{DandA}{J.S.Dowker and J.S.Apps \cqg{12}{95}{1363}.}
\ref{Watson1}{G.N.Watson {\it Theory of Bessel Functions} Cambridge
University Press, Cambridge, 1944.}
\ref{BGKE}{M.Bordag, B.Geyer, K.Kirsten and E.Elizalde, {\it Zeta function
determinant of the Laplace operator on the D-dimensional ball} UB-ECM-PF
95/10, hep-th /9505157.}
\ref{MandO}{W.Magnus and F.Oberhettinger {\it Formeln und S\"atze}
Springer-Verlag, Berlin, 1948.}
\ref{Olver}{F.W.J.Olver {\it Phil.Trans.Roy.Soc} {\bf A247} (1954) 328.}
\ref{Hurt}{N.E.Hurt {\it Geometric Quantization in action} Reidel,
Dordrecht, 1983.}
\ref{Esposito}{G.Esposito {\it Quantum Gravity, Quantum Cosmology and
Lorentzian Geometry}, Lecture Notes in Physics, Monographs, Vol. m12,
Springer-Verlag, Berlin 1994.}
\ref{Louko}{J.Louko \prD{38}{88}{478}.}
\ref{Schleich} {K.Schleich \prD{32}{85}{1989}.}
\ref{BEK}{M.Bordag, E.Elizalde and K.Kirsten {\it Heat kernel
coefficients of the Laplace operator on the D-dimensional ball}
UB-ECM-PF 95/3.}\ref{ELZ}{E.Elizalde, S.Leseduarte and S.Zerbini.}
\ref{BGV}{T.P.Branson, P.B.Gilkey and D.V.Vassilevich {\it The Asymptotics
of the Laplacian on a manifold with boundary} II, hep-th/9504029.}
\ref{Erdelyi}{A.Erdelyi,W.Magnus,F.Oberhettinger and F.G.Tricomi {\it
Higher Transcendental Functions} Vol.I McGraw-Hill, New York, 1953.}
\ref{Quine}{J.R.Quine, S.H.Heydari and R.Y.Song \tams{338}{93}{213}.}
\ref{Dikii}{L.A.Dikii {\it Usp. Mat. Nauk.} {\bf13} (1958) 111.}
\end{putreferences}
\bye